\documentclass[letterpaper, 10 pt, conference]{IEEEtran}  

\usepackage{graphics} 
\usepackage{epsfig} 
\usepackage{mathptmx} 
\usepackage{amsmath,amsfonts}
\usepackage{algorithmic}
\usepackage{algorithm}
\usepackage{array}
\usepackage{textcomp}
\usepackage{stfloats}
\usepackage{url}
\usepackage{verbatim}
\usepackage{graphicx}
\usepackage{cite}
\usepackage{times} 
\usepackage{textcomp}
\usepackage{amssymb}  
\usepackage{siunitx}
\usepackage[english]{babel}
\usepackage[autostyle]{csquotes}
\usepackage{booktabs}
\usepackage{multirow}
\usepackage{hyperref}
\usepackage{wrapfig}
\usepackage{fancyhdr}
\usepackage{enumitem}

\hypersetup{
    colorlinks=true,
    linkcolor=black,
    filecolor=magenta,      
    urlcolor=blue,
    pdfpagemode=FullScreen,
    citecolor=black,
    }

\title{\LARGE \bf
Vibration Compensation of Delta 3D Printer with Position-varying Dynamics using Filtered B-Splines
}

\author{Nosakhare Edoimioya, Cheng-Hao Chou, Chinedum E. Okwudire
}

\begin{document}

\maketitle
\thispagestyle{plain}
\pagestyle{plain}

\begin{abstract}

The delta robot is becoming a popular choice for the mechanical design of fused filament fabrication 3D printers because it can reach higher speeds than traditional serial-axis designs. Like serial 3D printers, delta printers suffer from undesirable vibration at high speeds which degrades the quality of fabricated parts. This undesirable vibration has been suppressed in serial printers using linear model-inversion feedforward control methods like the filtered B-splines (FBS) approach. However, techniques like the FBS approach are computationally challenging to implement on delta 3D printers because of their coupled, position-dependent dynamics. In this paper, we propose a methodology to address the computational bottlenecks by (1) parameterizing the position-dependent portions of the dynamics offline to enable efficient online model generation, (2) computing real-time models at sampled points (instead of every point) along the given trajectory, and (3) employing QR factorization to reduce the number of floating-point arithmetic operations associated with matrix inversion. In simulations, we report a computation time reduction of up to 23x using the proposed method, while maintaining high accuracy, when compared to a controller using the computationally expensive exact LPV model. In experiments, we demonstrate significant quality improvements on parts printed at various positions on the delta 3D printer using our proposed controller compared to a baseline alternative, which uses an LTI model from one position. Through acceleration measurements during printing, we also show that the print quality boost of the proposed controller is due to vibration reductions of more than 20\% when compared to the baseline controller.

\end{abstract}

\section{INTRODUCTION}

The delta robot is a high-speed, parallel-axis manipulator \cite{miller1995}, which makes it a promising candidate for increasing throughput in additive manufacturing. However, delta robots suffer from vibration errors that are a result of structural flexibilities in their kinematic chain \cite{zakhorov2022}; such vibration errors can adversely impact the quality of 3D printed parts. Unfortunately, delta 3D printers have not benefited from the model-based, feedforward control techniques that were recently used to suppress vibration on serial-axis 3D printers \cite{ramani2020,duan2018,kim2020,okwudire2018} because of the difficulty modeling and controlling the delta's coupled, nonlinear dynamics. These control techniques have resulted in up to 2x productivity increase without sacrificing accuracy on serial-axis printers \cite{okwudire2018}. This paper aims to address the challenges that prevent application of these control techniques on the delta 3D printer.

Previous work on modeling and controlling delta robots has largely been focused on rotary-joint delta robots, which are actuated with servo motors \cite{codourey1998,angel2018,boudjedir2018,ramirez-neria2015,castaneda2015,escorcia-hernandez2019,su2006,chiacchio1993,zhou2015,zhiyong2004}. (Most commercial delta 3D printers are prismatic-joint delta robots and typically use stepper motors). For servo motor delta robots, several methods have been studied--most of which rely on state measurements to estimate servo errors for accurate compensation. A PD or PID controller is usually a key element of these compensation methods. However, since standalone PD/PID controllers do not consider the dynamic coupling of delta robots, their performance is affected by the force disturbance inputs from other kinematic chains. To address this issue, Codourey \cite{codourey1998} combined a lumped model of the delta manipulator with a PD regulator in a computed torque (CT) control implementation to improve the tracking error performance in pick-and-place tasks when compared with a standalone PD regulator. Similarly, Angel and Viola \cite{angel2018} proposed a fractional PID controller combined with a CT controller. However, CT controllers need to have complete knowledge of the robot's dynamics, which can be challenging to obtain efficiently \cite{edoimioyaTASE2022}, and are sensitive to uncertainties and disturbance inputs. For example, in \cite{codourey1998}, workspace accelerations, which are necessary to calculate torques, are computed as second derivatives of the direct-geometric model of the robot (i.e., functions of joint positions). These calculations can be problematic when there is noise or other inaccuracies in the measurements. Perhaps this explains why no experiments that implement the controller on hardware are presented in \cite{angel2018} (only simulations). These challenges have led to the development of other techniques centered on servo error estimation and disturbance rejection \cite{boudjedir2018,ramirez-neria2015,castaneda2015,escorcia-hernandez2019,su2006,chiacchio1993}. These include methods like changing the PD gains online as a function of servo error estimates \cite{boudjedir2018}, disturbance rejection in the feedback loop using linear disturbance observers \cite{ramirez-neria2015,castaneda2015}, injecting inputs learned by a neural network to compensate errors that the PD controller does not reject \cite{escorcia-hernandez2019}, and using synchronization control strategies to reject coupling disturbances in each actuator from the other actuators \cite{su2006,chiacchio1993}. Furthermore, other approaches focus on tuning trajectory-dependent PID controller gains offline to minimize errors along a desired path that is known a priori \cite{zhou2015,zhiyong2004}. These PID gains provide reasonable tracking performance along the trajectory but require a priori knowledge of the entire trajectory. 

The central theme of all the methods discussed above is a dependence on feedback regulation to ease the difficulty of efficiently and accurately modeling the delta robot. However, most delta 3D printers cannot benefit from those methods because they do not have sensors for feedback control. Inspired by the feedforward CT controllers in \cite{codourey1998} and \cite{angel2018}, we recently proposed an efficient framework to obtain linear parameter-varying (LPV) models of prismatic-joint delta robots commonly used in 3D printers \cite{edoimioyaTASE2022}. The framework uses receptance coupling to split the full model of the delta robot into sub-models that can be independently identified using empirical measurements and analytical derivations. We demonstrated that the model accurately captures dynamic variation across different locations in the robot's task space. Accordingly, such a model allows a number of feedforward linear model-based vibration control techniques to be applied to the delta 3D printer. Among those is a class of methods known as model-inversion \cite{zundert2018}, which compensate vibration errors by using the inverse of the system's dynamics to pre-filter motion commands. Unlike other feedforward control methods such as smooth command generation \cite{erkorkmaz2001,tajima2022} and input shaping \cite{singhouse2009}, model-inversion does not introduce time delays \cite{ramani2017} and can theoretically lead to perfect compensation \cite{zundert2018}. In practice, perfect compensation is difficult to achieve due to unmodeled errors \cite{ramani2020} and the prevalence of nonminimum phase zeros, which can cause oscillatory or unbounded control commands. Nevertheless, several approximate model-inversion controllers have been employed in the literature \cite{zundert2018,rigney2009,clayton2009}. Of the available methods (as reviewed in \cite{zundert2018}), the filtered basis functions (FBF) approach has been shown to be versatile, compared to others, regarding its applicability to any linear system dynamics \cite{ramani2020,duan2018,okwudire2016,ramani2017,kasemsinsup2017,edoimioya2021}. The FBF approach expresses motion commands as a linear combination of basis functions, forward filters the basis functions using the system's dynamics, and calculates the coefficients of the basis functions that minimize motion errors. A version of FBF commonly used for controlling manufacturing machines is the filtered B-splines (FBS) method \cite{ramani2020,duan2018,kim2020,okwudire2018,okwudire2016,edoimioya2021}, where B-splines are selected as the basis functions because they are amenable to the lengthy motion trajectories common in manufacturing. FBS is implemented in real-time by sequentially processing small windows of the entire trajectory \cite{duan2018}. 

FBS has been implemented on serial-axis 3D printers, which are modeled as linear time-invariant (LTI) systems \cite{ramani2020,duan2018,kim2020,okwudire2018}, as well as a parallel-axis LPV 3D printer--the H-frame 3D printer \cite{edoimioya2021}. In the LTI (i.e., standard) implementation of FBS, B-splines are filtered and inverted offline to enable fast computation of their coefficients online. For the LPV system in \cite{edoimioya2021}, the authors model motion errors as a linear relationship between the $x$- and $y$-axis (and their LTI models). Furthermore, they approximate the dynamics as decoupled, resulting in independent computation of B-spline coefficients for each axis. Using these approximations, the B-splines can also be filtered and inverted offline. However, the delta 3D printer LPV model cannot be decoupled. Hence, its model needs to be recomputed at each new position, rendering real-time control with FBS computationally challenging. One must compute the new model, use it to filter the B-splines, and invert the filtered B-splines at every position along the trajectory.

This paper aims to address the computational challenges that hinder application of FBS on delta 3D printers by making the following contributions:
\begin{enumerate}
    \item We parameterize expressions of the delta's transfer functions offline, which leads to fast computation of the model online.
    \item We select one position per window of motion trajectory points as the position at which a model used to control all points in the window is generated. This choice leads to faster computation and lower memory allocation. We also propose a method to preserve continuity in the controller's prediction of output trajectories when the model switches between windows.
    \item We calculate the B-spline coefficients using QR factorization instead of pseudoinversion, leading to faster computations.
\end{enumerate}
The techniques we develop extend the standard FBS controller to the delta 3D printer. Furthermore, the proposed controller is shown to be effective at improving the quality of printed parts on a commercial delta 3D printer.

The rest of the paper is as follows: Section \ref{sec:dynamic_model} provides a recap of the dynamic model developed in \cite{edoimioyaTASE2022}; Section \ref{sec:proposed_controller} gives an overview of the standard FBS approach and describes the extensions we propose to enable real-time control of the delta 3D printer; Section \ref{sec:simulations_and_experiments} validates our proposed approach through simulations and experiments, illustrating the effectiveness of the proposed controller to improve print quality relative to a standard FBS controller; and Section \ref{sec:conclusions} concludes the paper, summarizing key insights. 

\section{OVERVIEW OF LPV MODEL OF DELTA 3D PRINTER}\label{sec:dynamic_model}

\subsection{Description of the delta 3D printer}

\begin{figure}[]
	\centering		\includegraphics[width=.48\textwidth]{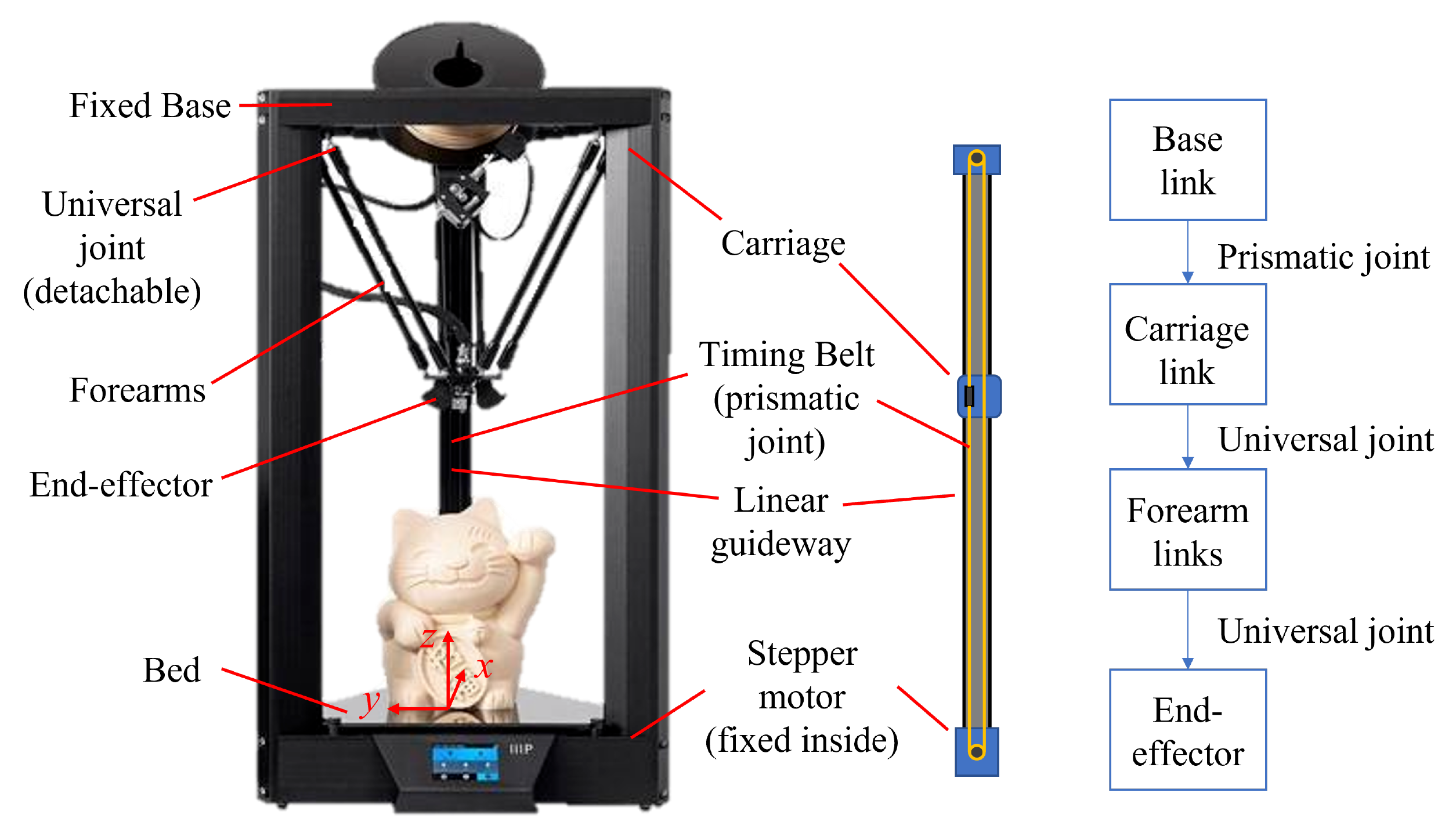}
    \caption{From left to right: A commercial delta 3D Printer (Monoprice Delta Pro) with labeled components, a schematic of the belt-driven carriage system, and the delta manipulator configuration showing the connections between joints and links. The print volume dimensions are 270 $\times$ 270 $\times$ 300 mm.}
    \label{fig:delta_schematic}
\end{figure}

As depicted in Fig. \ref{fig:delta_schematic}, three pairs of forearms on the delta 3D printer are connected to a carriage on one end and an end-effector on the other end. The carriages translate vertically to position one end of the forearms, which are allowed to rotate freely about universal joints. Each carriage moves on linear guideways and is mounted to a timing belt, which is, in turn, connected to a base-mounted stepper motor via a motor pulley--forming the prismatic joint. The relative position of each carriage (i.e., the joint space) determine the Cartesian position of the 3-DOF end-effector (i.e., the task space), which holds a nozzle that deposits melted filament onto a stationary bed. The parallelogram formed by each pair of forearms guarantees that the end-effector and fixed base remain co-planar.

\subsection{Linear parameter-varying model}

\begin{figure}[]
	\centering
    \includegraphics[width=.30\textwidth]{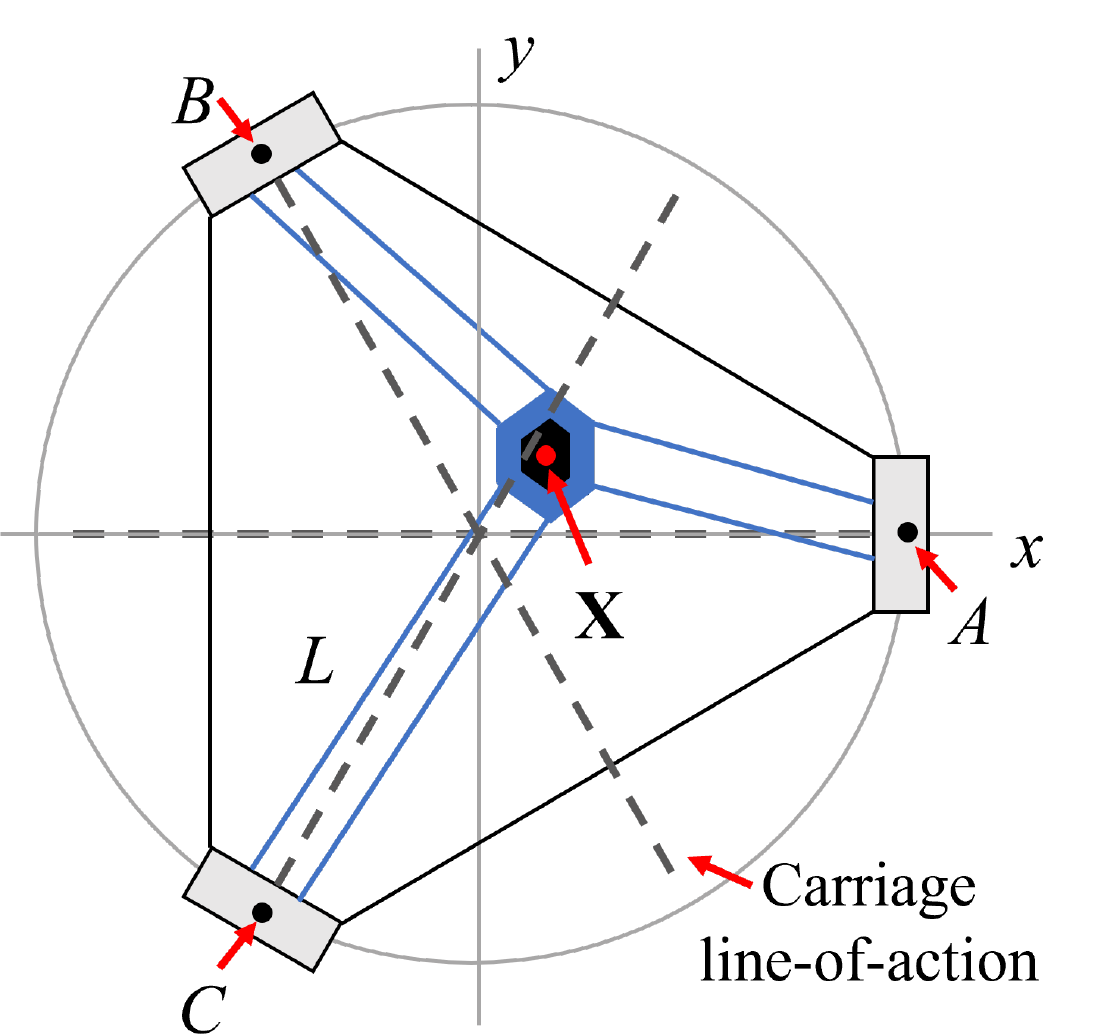}
    \caption{Overhead view of the delta 3D printer showing the $(x,y)$-coordinate locations of carriages $A$, $B$, and $C$, the end-effector's position in task space $\mathbf{X}$, and the length of the forearms $L$. End-effector motion along a carriage's line-of-action results in significant change to the carriage dynamics.}
    \label{fig:delta_overhead_view}
\end{figure}

The carriage output positions of the delta 3D printer, $q_{i}$, are a function of two inputs: (a) the commanded position of each carriage $q_{d_i}$ and (b) the forces $F_{q_i}$ imposed on each carriage due to the dynamics of the forearms and end-effector, where $i \in \{A,B,C\}$ denotes the carriages labeled $A$, $B$, and $C$ (see Fig. \ref{fig:delta_overhead_view}). Hence, the carriage output dynamics are given by
\begin{equation}
    q_{i}(s) = G_{q_{d}}(s)q_{d_i}(s) + G_{Fq}(s)F_{q_i}(\mathbf{X},s) \label{eq:output_q}
\end{equation}
where $s$ is the Laplace variable, $G_{q_{d}}(s)$ and $G_{Fq}(s)$ are LTI SISO systems representing the carriage position to position transfer function (TF) and the external force to carriage position TF, respectively, and $\mathbf{X}= [x \quad y \quad z]^T$ is the end-effector's position in the task space coordinates.

The coefficients of $G_{q_{d}}$ and $G_{Fq}$ can be identified from measurements on the printer and the parameters of the external forces $F_{q_i}$, which are modeled analytically, are also identified from measurements and least squares estimation techniques in \cite{edoimioyaTASE2022}. The analytical model of $F_{q_i}$ is obtained by considering the inertial dynamics of the end-effector in the task space, which are transformed into joint dynamics using the Jacobian transpose matrix. A detailed formulation of this process is given in \cite{edoimioyaTASE2022}. It transforms Eq. \eqref{eq:output_q} into the following expression:
\begin{equation}
    \mathbf{q}(s) = \mathbf{G}_{q_d}(s)\mathbf{q}_{d}(s) + \mathbf{G}_{Fq}(s)
    \begin{bmatrix}
        \bar{\mathbf{J}}_{A}^T\mathbf{P}_{A} \\
        \bar{\mathbf{J}}_{B}^T\mathbf{P}_{B} \\
        \bar{\mathbf{J}}_{C}^T\mathbf{P}_{C} 
    \end{bmatrix}
    \mathbf{W}(s)\bar{\mathbf{J}}\mathbf{q}(s)
    \label{eq:ouputQ}
\end{equation}
where $\mathbf{G}_{q_d}$ and $\mathbf{G}_{Fq}$ are $3\times 3$ diagonal matrices that contain $G_{q_{d}}$ and $G_{Fq}$, respectively, as the diagonal entries, $\mathbf{q} = [q_{A} \quad q_{B} \quad q_{C}]^T$ is the carriage output position vector, $\mathbf{q}_d = [q_{d_A} \quad q_{d_B} \quad q_{d_C}]^T$ is the desired position vector, $\mathbf{P}_{i} \in \mathbb{R}^{3\times 3}$ is the matrix representing the distribution of task space inertial forces associated with carriage $i$, $\bar{\mathbf{J}}_{i} \in \mathbb{R}^{3\times1}$, the column vector for carriage $i$ extracted from the linearized Jacobian matrix, denoted by $\bar{\textbf{J}}$ (see \cite{edoimioyaTASE2022} for details),
\begin{equation}
    \mathbf{W}(s) =
    \begin{bmatrix}
        w_{x}(s) & 0 & 0 \\
        0 & w_{y}(s) & 0 \\
        0 & 0 & w_{z}(s) 
    \end{bmatrix},
    \label{eq:W_defined}
\end{equation}
and $w_{x}$, $w_{y}$, and $w_{z}$ are the flexible inertial dynamics of the end-effector in the $x$-, $y$-, and $z$-axis directions, respectively. The model can be expressed simply as
\begin{equation}
    \mathbf{q}(s) = \mathbf{G}(s)\mathbf{q}_{d}(s)
    \label{eq:fullTF}
\end{equation}
where
\begin{equation}
    \mathbf{G}(s) = \Big[\mathbf{I} - \mathbf{G}_{Fq}(s)
    \begin{bmatrix}
        \bar{\mathbf{J}}_{A}^T\mathbf{P}_{A} \\
        \bar{\mathbf{J}}_{B}^T\mathbf{P}_{B} \\
        \bar{\mathbf{J}}_{C}^T\mathbf{P}_{C} 
    \end{bmatrix}
    \mathbf{W}(s)\bar{\mathbf{J}}\Big]^{-1}\mathbf{G}_{q_d}(s),
\end{equation}
yielding a linear parameter-varying (LPV) model of the delta 3D printer that can be used for linear model-inversion feedforward control. Since there are no position sensors, we assume the parameters of $\mathbf{G}(s)$ can be computed using the desired configuration instead of the output configuration, e.g., $\mathbf{X}_{d}$ instead of $\mathbf{X}$ (see Fig. \ref{fig:LPFBS_flowchart}), which was a reasonable assumption in previous work \cite{edoimioya2021}.




\section{FEEDFORWARD CONTROL WITH FILTERED B-SPLINES}\label{sec:proposed_controller}

\begin{figure*}
    \centering
    \includegraphics[width=0.8\textwidth]{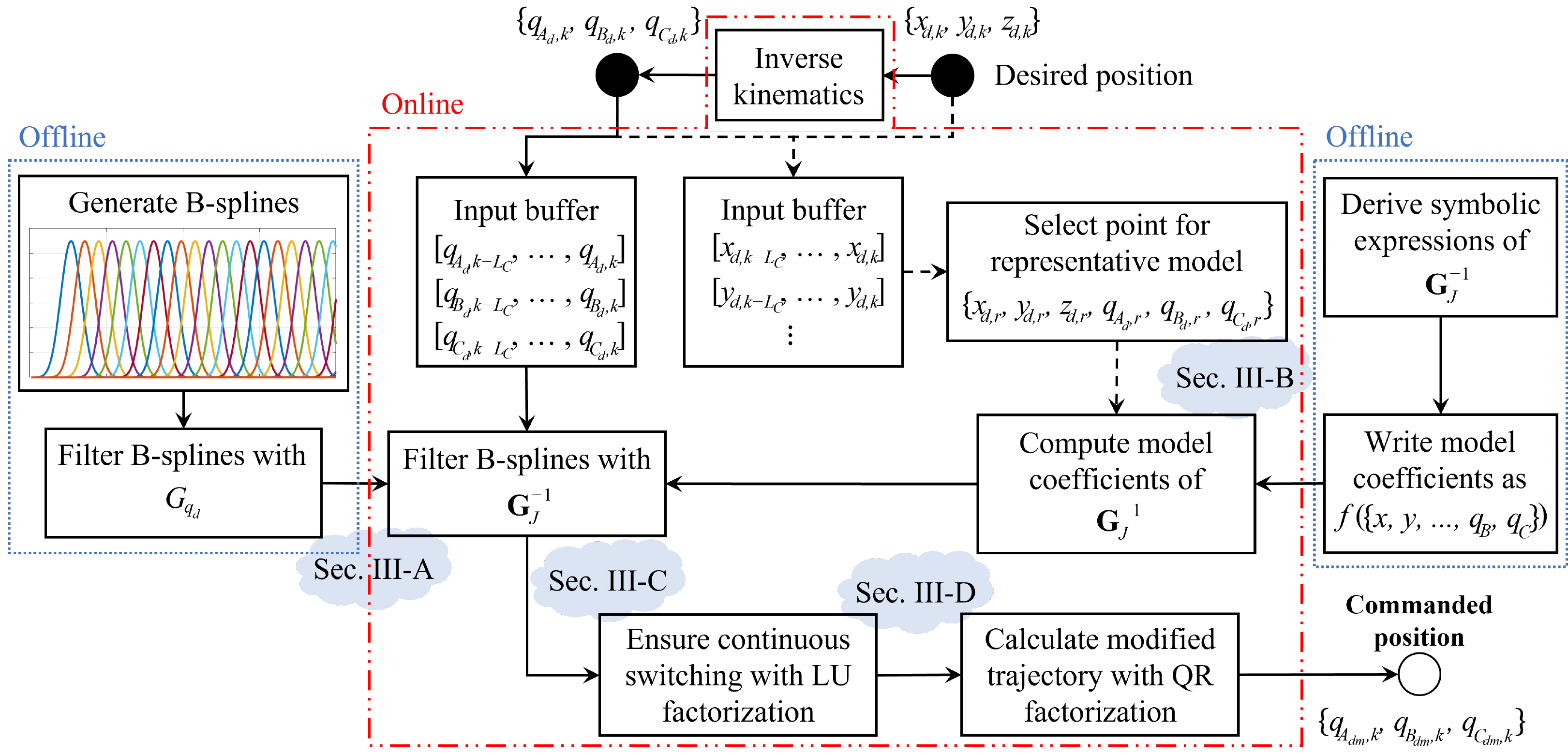}
    \caption{Flowchart of FBS implementation on delta 3D printer. First, the B-splines are generated offline and filtered with the carriage-only position-to-position dynamics as described Sec. \ref{subsec:FBS_overview} (left side). Online, $L_{C}$ points from the desired Cartesian and joint coordinates are buffered for a new window and a representative configuration is selected from the buffer--the median configuration in this paper. Then, the representative configuration $\{x_{d,r},...q_{C_{d},r}\}$ is used to compute the transfer function model coefficients of $\mathbf{G}_{J}^{-1}$ (see Sec. \ref{subsec:parameterization_and_filtering}). This transfer function is then used to filter the offline B-splines. Finally, we compute the approximate B-spline coefficients from the previous window to maintain continuity during switching (Sec. \ref{subsec:switching}) and calculate the modified trajectory (Sec. \ref{subsec:QR_decomposition}).}
    \label{fig:LPFBS_flowchart}
\end{figure*}

In this section, we give an overview of the standard FBS approach in Section \ref{subsec:FBS_overview}. Then, we discuss the process of extending FBS by: describing the selection of a parameterized model to filter the B-splines in Section \ref{subsec:parameterization_and_filtering}, explaining how continuity is preserved when switching models between windows in Section \ref{subsec:switching}, and describing how the motion command is generated using LU and QR factorization in Section \ref{subsec:QR_decomposition}. As a visual aid, the reader can follow a flowchart of the process of generating optimal motion commands in Fig. \ref{fig:LPFBS_flowchart}.

\subsection{Overview of the standard FBS approach}\label{subsec:FBS_overview}
The FBS approach (as presented in \cite{duan2018}) controls the lifted system representation (LSR) of $\mathbf{G}(s)$ with a feedforward controller. (See Appendix A for details on the LSR). Let $\mathbf{q}_{i_d} = [q_{i_d}(t_{0})\hspace{0.5em} q_{i_d}(t_{1})\hspace{0.5em} \cdots\hspace{0.5em} q_{i_d}(t_{E})]^{T}$ represent the entire $E+1$ discrete time steps of the desired trajectory of carriage $i$, which are processed in sliding windows. Assume that time step $k$ marks the beginning of the current window and that the unknown modified motion command $\mathbf{q}_{i_{dm},\text{C}} = [q_{i_{dm}}(t_k)\hspace{0.5em} q_{i_{dm}}(t_{k+1}) \hspace{0.5em} \cdots\hspace{0.5em} q_{i_{dm}}(t_{k+L_{C}})]^{T}$ is parameterized using B-splines such that
\begin{equation}
    \begin{bmatrix}
       q_{i_{dm}}(t_{k}) \\
       q_{i_{dm}}(t_{k+1}) \\
       \vdots \\
       q_{i_{dm}}(t_{k+L_{C}})
    \end{bmatrix}
    = \underbrace{
    \begin{bmatrix}
       \phi_{m,m}(t_{k}) & \cdots & \phi_{m+n,m}(t_{k}) \\
       \phi_{m,m}(t_{k+1}) & \cdots & \phi_{m+n,m}(t_{k+1}) \\
       \vdots & \ddots & \vdots \\
       \phi_{m,m}(t_{k+L_{C}}) & \cdots & \phi_{m+n,m}(t_{k+L_{C}}) \\
    \end{bmatrix}
    }_{\Phi}
    \underbrace{
    \begin{bmatrix}
       p_{i,m} \\
       \vdots \\
       p_{i,m+n}
    \end{bmatrix}
    }_{\mathbf{p}_{i,\text{C}}}
    \label{eq:fbs_parameterization}
\end{equation}
where the (non-italicized) subscript C denotes the current window, $L_C$ is the number of trajectory points considered for each window, $\Phi$ is the open-ended B-spline basis functions matrix of degree $m$ \cite{duan2018}, $\phi_{j,m}(t)$ are real-valued basis functions \cite{nurbsbook}, $j=m,m+1,...,m+n$, $\mathbf{p}_{i,\text{C}}$ is a vector of $n+1$ unknown coefficients (or control points), $t_{k} = kT_{s}$ is the current time, and $T_{s}$ is the sampling time (see \cite{duan2018} for more details). To capture the coupling between carriages, we define $\mathbf{q}_{d,\text{C}} = [\mathbf{q}_{A_d,\text{C}}^T \hspace{0.5em}\mathbf{q}_{B_d,\text{C}}^T \hspace{0.5em}\mathbf{q}_{C_d,\text{C}}^T]^{T}$, such that
\begin{equation}
    \mathbf{q}_{dm,\text{C}}=
    \begin{bmatrix}
       \mathbf{q}_{A_{dm},\text{C}} \\
       \mathbf{q}_{B_{dm},\text{C}} \\
       \mathbf{q}_{C_{dm},\text{C}}
    \end{bmatrix}
    = \underbrace{
    \begin{bmatrix}
       \Phi & \mathbf{0} & \mathbf{0} \\
       \mathbf{0} & \Phi & \mathbf{0} \\
       \mathbf{0} & \mathbf{0} & \Phi \\
    \end{bmatrix}
    }_{\mathbf{N}_{\text{C}}}
    \underbrace{
    \begin{bmatrix}
       \mathbf{p}_{A,\text{C}} \\
       \mathbf{p}_{B,\text{C}} \\
       \mathbf{p}_{C,\text{C}}
    \end{bmatrix}
    }_{\mathbf{p}_{\text{C}}}.
    \label{eq:bspline_parameterization}
\end{equation}
Our objective is to minimize the tracking error, defined as
\begin{multline}
    \bar{\mathbf{e}} = \mathbf{q}_d - \bar{\mathbf{N}}\bar{\mathbf{p}} \Leftrightarrow \\
    \begin{bmatrix}
       \bar{\mathbf{e}}_{\text{P}} \\
       \bar{\mathbf{e}}_{\text{C}} \\
       \bar{\mathbf{e}}_{\text{F}} 
    \end{bmatrix}
    =
    \begin{bmatrix}
       \mathbf{q}_{d,\text{P}} \\
       \mathbf{q}_{d,\text{C}} \\
       \mathbf{q}_{d,\text{F}} 
    \end{bmatrix}
    -
    \begin{bmatrix}
       \bar{\mathbf{N}}_{\text{P}} & \mathbf{0} & \mathbf{0} \\
       \bar{\mathbf{N}}_{\text{PC}} & \bar{\mathbf{N}}_{\text{C}} & \mathbf{0} \\
       \mathbf{0} & \bar{\mathbf{N}}_{\text{CF}} & \bar{\mathbf{N}}_{\text{F}}
    \end{bmatrix}
    \begin{bmatrix}
       \bar{\mathbf{p}}_{\text{P}} \\
       \bar{\mathbf{p}}_{\text{C}} \\
       \bar{\mathbf{p}}_{\text{F}} 
    \end{bmatrix}
    \label{eq:LPFBS_matrix_expansion}
\end{multline}
where subscripts P and F denote the past and future windows, respectively, 
\begin{equation}
    \mathbf{q}_{\text{C}} = \bar{\mathbf{N}}_{\text{C}}\bar{\mathbf{p}}_{\text{C}} + \bar{\mathbf{N}}_{\text{PC}}\bar{\mathbf{p}}_{\text{P}}
\end{equation}
represents the current output carriage motion,
and the bar on the matrices and vectors indicates that the impulse response of the transfer function used for filtering the B-splines is truncated \cite{duan2018}. Using local least squares, the optimal coefficients of the current window can be computed as 
\begin{align}
    \bar{\mathbf{p}}_{\text{C}}  & = (\bar{\mathbf{N}}_{\text{C}}^{T}\bar{\mathbf{N}}_{\text{C}})^{-1}\bar{\mathbf{N}}_{\text{C}}\Big(\mathbf{q}_{d,\text{C}} - \bar{\mathbf{N}}_{\text{PC}}\bar{\mathbf{p}}_{\text{P}}\Big)\\
    & = \bar{\mathbf{N}}_{\text{C}}^{\dagger}\Big(\mathbf{q}_{d,\text{C}} - \bar{\mathbf{N}}_{\text{PC}}\bar{\mathbf{p}}_{\text{P}}\Big) \label{eq:LPFBS_LTI_optimization}
\end{align}
where $\bar{\mathbf{p}}_{\text{P}}$ denotes the coefficients calculated in the previous window. Note that although $n$ coefficents are computed, only $n_{\text{up}}$ are updated in each window \cite{duan2018}. 

For an LTI system, $\mathbf{N}_{\text{C}}$ is pre-filtered and $\bar{\mathbf{N}}_{\text{PC}}$ and $\bar{\mathbf{N}}_{\text{C}}^{\dagger}$ are computed offline and stored for obtaining the optimal coefficients in every window using Eq. \eqref{eq:LPFBS_LTI_optimization}. However, for LPV systems, filtering and inverting the (large) B-splines matrix in real-time is computationally challenging for most hardware processors. Constrained by the system's computation and memory capabilities, the rest of this section proposes techniques to optimize the computation and memory resources required to apply FBS to the delta 3D printer without significantly sacrificing the achieved accuracy improvement.

\subsection{Selecting a parameterized model for B-splines filtering}\label{subsec:parameterization_and_filtering}

Consider the problem of filtering each column of $\mathbf{N}$ through $\mathbf{G}(s)$, reproduced below:

\begin{equation}
    \mathbf{G}(s) = \underbrace{\Big[\mathbf{I} - \mathbf{G}_{Fq}(s)
    \begin{bmatrix}
        \bar{\mathbf{J}}_{A}^T\mathbf{P}_{A} \\
        \bar{\mathbf{J}}_{B}^T\mathbf{P}_{B} \\
        \bar{\mathbf{J}}_{C}^T\mathbf{P}_{C} 
    \end{bmatrix}
    \mathbf{W}(s)\bar{\mathbf{J}}\Big]^{-1}}_{\mathbf{G}_{J}^{-1}(s)}\mathbf{G}_{q_d}(s).
\end{equation}
Note that $\mathbf{G}_{J}^{-1}\in \mathbb{R}^{3\times 3}$ depends on the configuration through the Jacobian matrix $\bar{\textbf{J}}$, while $\mathbf{G}_{q_d}$ is not position dependent. Hence, we can derive symbolic expressions of each transfer function in $\mathbf{G}_{J}^{-1}$ as functions of position. This derivation leads to symbolic transfer functions of the form
\begin{equation}
    G_{J,AA}^{-1} = \frac{b_{AA}(x,y,z,q_{A},q_{B},q_{C})}{a(x,y,z,q_{A},q_{B},q_{C})}
    \label{eq:parameterized_Gjinv}
\end{equation}
where $b_{AA}(\cdot)$ and $a(\cdot)$ are the numerator and denominator of the transfer function, respectively, and the subscript \enquote{$AA$} denotes values pertaining to the $A$-to-$A$ carriage position dynamics. The other 8 transfer functions ($G_{J,BA}^{-1}$, $G_{J,CA}^{-1}$, $G_{J,AB}^{-1}$, and so on) can be expressed similarly with $b_{BA}(\cdot)$, $b_{CA}(\cdot)$, $b_{AB}(\cdot)$, and so on. Note that all transfer functions share the same denominator  $a(\cdot)$. These parameterized transfer functions enable fast computations of the coefficients of $\mathbf{G}_{J}^{-1}$ during real-time control by simply substituting the corresponding values of $x$, $y$, $z$, $q_{A}$, $q_{B}$, and $q_{C}$ into the symbolic expressions. Furthermore, we can pre-filter $\mathbf{N}$ with $\mathbf{G}_{q_d}$ offline to obtain $\bar{\mathbf{N}}_{q_{d}}$. Then, for each window of trajectory points processed, we filter $\bar{\mathbf{N}}_{q_{d}}$ with the transfer functions in Eq. \eqref{eq:parameterized_Gjinv} to obtain 
\begin{equation}
    \begin{bmatrix}
       \bar{\mathbf{N}}_{\text{C}}\\
       \bar{\mathbf{N}}_{\text{CF}}
    \end{bmatrix} = 
    \begin{bmatrix}
        \vspace{0.5em}
       \begin{bmatrix}\bar{\mathbf{N}}_{\text{C}_{AA}} \\ \bar{\mathbf{N}}_{\text{CF}_{AA}}\end{bmatrix}& \begin{bmatrix}\bar{\mathbf{N}}_{\text{C}_{BA}} \\ \bar{\mathbf{N}}_{\text{CF}_{BA}}\end{bmatrix}& \begin{bmatrix}\bar{\mathbf{N}}_{\text{C}_{CA}} \\ \bar{\mathbf{N}}_{\text{CF}_{CA}}\end{bmatrix} \\
       
       \vspace{0.5em}
       
       \begin{bmatrix}\bar{\mathbf{N}}_{\text{C}_{AB}} \\ \bar{\mathbf{N}}_{\text{CF}_{AB}}\end{bmatrix}& \begin{bmatrix}\bar{\mathbf{N}}_{\text{C}_{BB}} \\ \bar{\mathbf{N}}_{\text{CF}_{BB}}\end{bmatrix}&        \begin{bmatrix}\bar{\mathbf{N}}_{\text{C}_{CB}} \\ \bar{\mathbf{N}}_{\text{CF}_{CB}} \end{bmatrix} \\
       
       \begin{bmatrix}\bar{\mathbf{N}}_{\text{C}_{AC}} \\ \bar{\mathbf{N}}_{\text{CF}_{AC}} \end{bmatrix}& \begin{bmatrix}\bar{\mathbf{N}}_{\text{C}_{BC}} \\ \bar{\mathbf{N}}_{\text{CF}_{BC}}\end{bmatrix} & \begin{bmatrix}\bar{\mathbf{N}}_{\text{C}_{CC}} \\ \bar{\mathbf{N}}_{\text{CF}_{CC}}\end{bmatrix}
    \end{bmatrix}
\end{equation}
where $[\bar{\mathbf{N}}_{\text{C}_{AA}}^{T} \quad \bar{\mathbf{N}}_{\text{CF}_{AA}}^{T}]^T$ is the result of filtering the columns of $\bar{\mathbf{N}}_{q_{d}}$ through $G_{J,AA}^{-1}$, and so on for the other blocks of the matrix.

In practice, the current and future windows are overlapped for continuity during computation but only $L_{C}$ points are updated during each sequence \cite{duan2018}. Therefore, each overlapped window has $2L_{C}$ trajectory points, meaning that the time complexity for computing the transfer function coefficients is $O(2L_{C})$ (assuming parallel computation) and the space complexity is $O(2L_{c}u_{a})$, where $u_{a}$ is the order of the transfer functions. Some computers may not have enough processing power to complete these calculations while maintaining real-time printing--especially the smaller micro-processors commonly used by 3D printer manufacturers. Additionally, allocating the memory resources required to store the coefficients may limit the computer's ability to allocate quick-access memory to other important functions like storing the print trajectory. 

To prevent such deleterious effects, we select one point from each window (of the first $L_{C}$ points) where a transfer function is computed, which reduces the time and space complexity to $O(1)$ and $O(u_{a})$, respectively. This trade-off is reasonable because $L_{C}$ generally represents a small distance where the dynamics do not change significantly. For example, $L_{C}$ typically ranges from 100-200 points, which represents 100 to 200 ms for a standard sampling interval of 1 ms. For most practical applications, the 3D printer will not cover large enough distances in $\le$200 ms to create significant dynamic variation. For our implementation in Section \ref{sec:simulations_and_experiments}, we select the median point (i.e., the point in the middle of the window) as the representative point. One can select other points such as the mean point (i.e., the average point in the window for each configuration variable, considered independently) or the point with the minimum total Euclidean distance from all the other points in the same window. In simulations, we found that the tracking accuracy is not significantly different when any reasonable central point is selected. In Section \ref{sec:simulations_and_experiments}, we demonstrate that selecting one point in each window does not significantly degrade the accuracy of the controller through simulations and experiments.

\subsection{Smoothly switching models between windows}\label{subsec:switching}

One drawback to selecting a different model for each window is that switching models can lead to discontinuities in the controller's predicted output trajectories. In the standard FBS approach, continuity is preserved by using the same LTI model to predict the trajectory for every window \cite{duan2018}. 
\begin{figure}[]
	\centering
    \includegraphics[width=.45\textwidth]{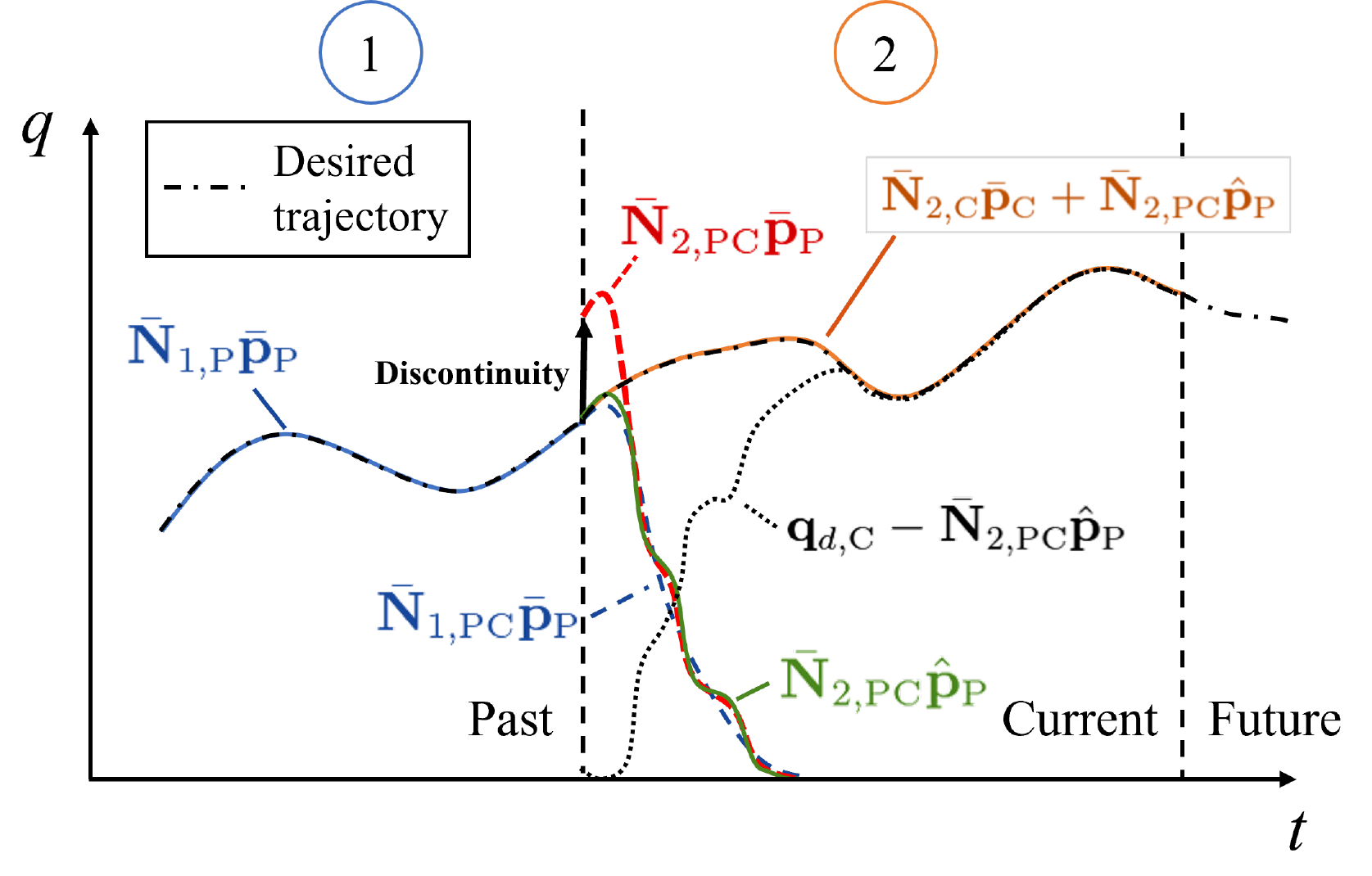}
    \caption{Illustration of the switching compensation technique to maintain continuity described in Sec. \ref{subsec:switching}. The B-spline coefficients (or control points) from the previous window $\bar{\mathbf{p}}_{\text{P}}$ are approximated as $\hat{\mathbf{p}}_{\text{P}}$ to maintain continuity when the model is switched from model 1 to model 2. Note that $\bar{\mathbf{N}}_{1,\text{PC}}\bar{\mathbf{p}}_{\text{P}}$ does not have the correct dynamics for the current window and $\bar{\mathbf{N}}_{2,\text{PC}}\bar{\mathbf{p}}_{\text{P}}$ creates a discontinuity at the window boundary. The difference between the desired trajectory and the approximate residual motion is also shown as $\mathbf{q}_{d,\text{C}} - \bar{\mathbf{N}}_{\text{PC}}\hat{\mathbf{p}}_{\text{P}}$.}
    \label{fig:switching_compensation}
\end{figure}
To demonstrate what happens in the LPV case, suppose we used model 1 for the past window and update it to model 2 for the current window as shown in Fig. \ref{fig:switching_compensation}. When we switch from model 1 to model 2, note that the prediction of the output trajectory in the current window will be different depending on if we use model 1 (i.e., $\bar{\mathbf{N}}_{1,\text{PC}}\bar{\mathbf{p}}_{\text{P}}$) or model 2 (i.e., $\bar{\mathbf{N}}_{2,\text{PC}}\bar{\mathbf{p}}_{\text{P}}$) for the prediction. Since model 2 captures the dynamics in the current window more accurately than model 1,  $\bar{\mathbf{N}}_{2,\text{PC}}$ should be used for the prediction. However, using $\bar{\mathbf{N}}_{2,\text{PC}}$ may result in a discontinuity in the prediction of the machine's motion at the point where the window changes because the previous window's control points, $\bar{\mathbf{p}}_{\text{P}}$, were computed using model 1. 

To resolve this discrepancy, we approximate the prediction by generating a set of approximate control points that ensure continuity with the output from the past window. The approximate control points are selected to minimize the difference between the new prediction and the (potentially) discontinuous prediction while preserving continuity. We write the optimization problem as
\begin{equation} \label{eq:int_coefficients_optimization}
    \begin{split}
        \hat{\mathbf{p}}_{\text{P}} = \arg \min_{\hat{\mathbf{p}}_{\text{P}}} & \quad \|\bar{\mathbf{N}}_{2,\text{PC}}\hat{\mathbf{p}}_{\text{P}} - \bar{\mathbf{N}}_{2,\text{PC}}\bar{\mathbf{p}}_{\text{P}}\|_{2}^{2} \\
        s.t. & \quad \bar{N}_{2,\text{PC}}^{T}(t_{k})\hat{\mathbf{p}}_{\text{P}} = \bar{N}_{1,\text{PC}}^{T}(t_{k})\bar{\mathbf{p}}_{\text{P}} \\
         & \quad \bar{N'}_{2,\text{PC}}^{T}(t_{k})\hat{\mathbf{p}}_{\text{P}} = \bar{N'}_{1,\text{PC}}^{T}(t_{k})\bar{\mathbf{p}}_{\text{P}}
    \end{split}
\end{equation}
where $\bar{N}_{1,\text{PC}}^{T}(t_{k})$ and $\bar{N}_{2,\text{PC}}^{T}(t_{k})$ are the first rows of $\bar{\mathbf{N}}_{1,\text{PC}}$ and $\bar{\mathbf{N}}_{2,\text{PC}}$ in the window, respectively, $\bar{N'}_{1,\text{PC}}^{T}(t_{k})$ and $\bar{N'}_{2,\text{PC}}^{T}(t_{k})$ are the first rows of $\bar{\mathbf{N}'}_{1,\text{PC}}$ and $\bar{\mathbf{N}'}_{2,\text{PC}}$, respectively (which are the time derivatives of $\bar{\mathbf{N}}_{1,\text{PC}}$ and $\bar{\mathbf{N}}_{2,\text{PC}}$), and $\hat{\mathbf{p}}_{\text{P}}$ are the approximate control points. Note that the products 
\begin{equation}
    \bar{N}_{1,\text{PC}}^{T}(t_{k})\bar{\mathbf{p}}_{\text{P}} \quad \text{and} \quad \bar{N}_{2,\text{PC}}^{T}(t_{k})\hat{\mathbf{p}}_{\text{P}}
\end{equation}
represent positions at the window boundary, and 
\begin{equation}
    \bar{N'}_{1,\text{PC}}^{T}(t_{k})\bar{\mathbf{p}}_{\text{P}} \quad \text{and} \quad \bar{N'}_{2,\text{PC}}^{T}(t_{k})\hat{\mathbf{p}}_{\text{P}}
\end{equation}
and represent velocities at the boundary. Additional kinematic constraints, such as acceleration and jerk, can be included in the optimization problem from Eq. \eqref{eq:int_coefficients_optimization} by taking additional derivatives of the B-splines as described in \cite{nurbsbook} and \cite{duan2015IJAMT}. More kinematic constraints leads to smoother transitions when the dynamics change significantly or when the window size is large. In our simulations of the machine used in Section \ref{sec:simulations_and_experiments}, we found that position and velocity constraints led to similar tracking accuracy when compared to optimizing Eq. \eqref{eq:int_coefficients_optimization} with acceleration and jerk constraints. Hence, our implementation only uses the position and velocity constraints for Eq. \eqref{eq:int_coefficients_optimization}.

Using the approximate control points, the coefficients that minimize the tracking error in the current window are obtained by solving
\begin{multline} \label{eq:LPFBS_LTI_optimization_modified}
    \bar{\mathbf{p}}_{\text{C}} = \arg\min_{\bar{\mathbf{p}}_{\text{C}}}\Big[\Big((\mathbf{q}_{d,\text{C}} - \bar{\mathbf{N}}_{\text{PC}}\hat{\mathbf{p}}_{\text{P}}) - \bar{\mathbf{N}}_{\text{C}}\bar{\mathbf{p}}_{\text{C}}\Big)^{T}\\
    \Big((\mathbf{q}_{d,\text{C}} - \bar{\mathbf{N}}_{\text{PC}}\hat{\mathbf{p}}_{\text{P}}) - \bar{\mathbf{N}}_{\text{C}}\bar{\mathbf{p}}_{\text{C}}\Big)\Big].
\end{multline}

Equation \eqref{eq:LPFBS_LTI_optimization_modified} can be computationally expensive to solve in real-time for each window using the pseudoinverse, as done in Eq. \eqref{eq:LPFBS_LTI_optimization}. Similarly, solving the constrained optimization problem in Eq. \eqref{eq:int_coefficients_optimization} in real-time could be challenging. To speed up the computations, we employ the LU and QR factorization methods for solving Eqs. \eqref{eq:int_coefficients_optimization} and \eqref{eq:LPFBS_LTI_optimization_modified}, respectively, as discussed in the following subsection.

\subsection{Command generation with LU and QR factorization}\label{subsec:QR_decomposition}

The optimization problem in Eq. \eqref{eq:int_coefficients_optimization} can be solved with a number of gradient-based algorithms. For example, Matlab provides functions \textit{fmincon} and \textit{lsqlin} to solve constrained optimization problems. However, such algorithms may require a large number of iterations to converge to a solution, which can stall our controller. To circumvent this problem, we can solve the constrained least squares problem with LU factorization by leveraging properties of the filtered B-splines. To simplify notation, we define the following from Eq. \eqref{eq:int_coefficients_optimization}:
\begin{align}
    \mathbf{A} & =  \bar{\mathbf{N}}_{2,\text{PC}}, \quad \mathbf{b}  = \bar{\mathbf{N}}_{2,\text{PC}}\bar{\mathbf{p}}_{\text{P}} \\
    \mathbf{C} & = 
    \begin{bmatrix}
       \bar{N}_{2,\text{PC}}^{T}(t_{k}) \\
       \bar{N'}_{2,\text{PC}}^{T}(t_{k})
    \end{bmatrix}, \quad
    \mathbf{d} = 
    \begin{bmatrix}
       \bar{N}_{1,\text{PC}}^{T}(t_{k})\bar{\mathbf{p}}_{\text{P}}\\
       \bar{N'}_{1,\text{PC}}^{T}(t_{k})\bar{\mathbf{p}}_{\text{P}}
    \end{bmatrix}
\end{align}
Then, the problem can be written as
\begin{equation} 
    \begin{split}
        \hat{\mathbf{p}}_{\text{P}} = \arg \min_{\hat{\mathbf{p}}_{\text{P}}} & \quad \|\mathbf{A}\hat{\mathbf{p}}_{\text{P}} - \mathbf{b}\|_{2}^{2} \\
        s.t. & \quad \mathbf{C}\hat{\mathbf{p}}_{\text{P}} = \mathbf{d} \\
    \end{split}
\end{equation}
We make two assumptions:
\begin{enumerate}
    \item The stacked matrix 
    \begin{equation}
        \begin{bmatrix}
           \mathbf{A} \\
           \mathbf{C}
        \end{bmatrix}
    \end{equation}
    has linearly independent columns; and
    \item $\mathbf{C}$ has linearly independent rows.
\end{enumerate}
As discussed in \cite{ramani2017}, the filtered B-splines satisfy the above assumptions with high probability and, in the case they do not, the B-splines can be freely selected by the user to satisfy the assumptions. Then, we can construct the Lagrangian,
\begin{equation}
    \mathcal{L}(\hat{\mathbf{p}}_{\text{P}},\lambda) \triangleq \frac{1}{2}\|\mathbf{A}\hat{\mathbf{p}}_{\text{P}} - \mathbf{b}\|_{2}^{2} + \lambda^T(\mathbf{C}\hat{\mathbf{p}}_{\text{P}} - \mathbf{d}),
\end{equation}
where $\lambda$ is a set of Lagrange multipliers, and find where its partial derivatives equal zero to obtain the following linear system
\begin{equation} \label{eq:lagrange_equations}
    \begin{bmatrix}
       \mathbf{A}^T\mathbf{A} & \mathbf{C}^T \\
       \mathbf{C} & \mathbf{0}
    \end{bmatrix}
    \begin{bmatrix}
       \hat{\mathbf{p}}_{\text{P}} \\
       \lambda
    \end{bmatrix}
    =
    \begin{bmatrix}
        \mathbf{A}^T\mathbf{b} \\
        \mathbf{d}
    \end{bmatrix}.
\end{equation}
Note that the matrix 
\begin{equation}
    \begin{bmatrix}
       \mathbf{A}^T\mathbf{A} & \mathbf{C}^T \\
       \mathbf{C} & \mathbf{0}
    \end{bmatrix}
\end{equation}
is nonsingular when the above assumptions hold. Therefore, the linear equation given by Eq. \eqref{eq:lagrange_equations} can be efficiently solved with LU factorization \cite{matrix_computations}.

We also use QR factorization to efficiently compute the control points. Using the pseudoinverse to solve the optimization problem in Eq. \eqref{eq:LPFBS_LTI_optimization_modified} requires the following number of floating-point operations (flops) \cite{matrix_computations}:
\begin{align}
    \bar{\mathbf{N}}_{\text{C}}^{T}\bar{\mathbf{N}}_{\text{C}} \hspace{0.1em} &: \hspace{0.1em} L_{C}n^2 \hspace{0.5em} \text{flops} \\
    (\bar{\mathbf{N}}_{\text{C}}^{T}\bar{\mathbf{N}}_{\text{C}})^{-1} \hspace{0.1em} &: \hspace{0.1em} n^3 + L_{C}n^2 \hspace{0.5em} \text{flops} \\
    \bar{\mathbf{N}}_{\text{C}}\tilde{\mathbf{q}}_{d,\text{C}} \hspace{0.1em} &: \hspace{0.1em} n^3 + L_{C}n^2 + 2L_{C}n \hspace{0.5em} \text{flops} \\
    (\bar{\mathbf{N}}_{\text{C}}^{T}\bar{\mathbf{N}}_{\text{C}})^{-1}\Big(\bar{\mathbf{N}}_{\text{C}}\tilde{\mathbf{q}}_{d,\text{C}}\Big) \hspace{0.1em} &: \hspace{0.1em} n^3 + L_{C}n^2 + 4L_{C}n \hspace{0.5em} \text{flops}.
\end{align}
where $\tilde{\mathbf{q}}_{d,\text{C}} = \mathbf{q}_{d,\text{C}} - \bar{\mathbf{N}}_{\text{PC}}\hat{\mathbf{p}}_{\text{P}}$. By factoring 
\begin{equation}
    \bar{\mathbf{N}}_{\text{C}} = \mathbf{Q}\mathbf{R}
\end{equation}
with the modified Gram Schmidt algorithm \cite{numerical_analysis}, where $\mathbf{Q} \in \mathbb{R}^{L_{C}\times L_{C}}$ is an orthogonal matrix (i.e., $\mathbf{Q}^{T}\mathbf{Q} = \mathbf{I}$) and $\mathbf{R}\in \mathbb{R}^{L_{C}\times n}$ is an upper triangular matrix, the problem in Eq. \eqref{eq:LPFBS_LTI_optimization_modified} can be written as
\begin{equation}
    \mathbf{R}\bar{\mathbf{p}}_{c,\text{C}} = \mathbf{Q}^{T}\tilde{\mathbf{q}}_{d,\text{C}}
\end{equation}
which can be solved using backward substitution. The number of operations required using this method are
\begin{align}
    \bar{\mathbf{N}}_{\text{C}} = \mathbf{Q}\mathbf{R} \hspace{0.1em} &: \hspace{0.1em} L_{C}n^2 \hspace{0.5em} \text{flops} \\
    \mathbf{w} = \mathbf{Q}^{T}\tilde{\mathbf{q}}_{d,\text{C}} \hspace{0.1em} &: \hspace{0.1em} L_{C}n^2 + 2L_{C}n \hspace{0.5em} \text{flops} \\
    \mathbf{R}\bar{\mathbf{p}}_{c,\text{C}} = \mathbf{w} \hspace{0.1em} &: \hspace{0.1em} n^2 + L_{C}n^2 + 2L_{C}n \hspace{0.5em} \text{flops}.
\end{align}
Note that the QR factorization solution is more efficient to compute than the pseudoinverse.

\section{SIMULATION AND EXPERIMENTAL VALIDATION}\label{sec:simulations_and_experiments}
\subsection{Simulation validation}\label{subsec:simulations}
In this subsection, we focus on simulations of the MP Delta Pro 3D printer from Fig. \ref{fig:delta_schematic}. The simulation model is used to validate our assumptions about computation time and preserved accuracy of the controller proposed in Section \ref{sec:proposed_controller}. We evaluate the performance of the following controllers: 
\begin{enumerate}[label=(\alph*)]
    \item a controller where the transfer functions are computed in matrix form using Eq. \eqref{eq:fullTF} for all points in the window and the coefficients are calculated with the pseudoinverse;
    \item a controller that is the same as controller (a), except the transfer functions are computed using the parameterized model (Sec. \ref{subsec:parameterization_and_filtering});
    \item a controller that is the same as controller (b), except the transfer functions are computed for only one point in the window--\textit{without} the switching compensation discussed in Sec. \ref{subsec:switching};
    \item a controller that is the same as controller (c), except \textit{with} the switching compensation; and
    \item a controller that is the same as controller (d), except the coefficients are calculated with QR factorization (our proposed controller).
\end{enumerate}

By adding the proposed modifications separately, we can distinguish the computational efficiency and accuracy effects of each modification. Each controller's performance is compared to a baseline controller--the standard FBS controller using an LTI model for the carriages at $(x,y) = (0,0)$ mm. Without a model for the position-dependent dynamics, the control designer must use one LTI model for FBS and a reasonable choice is the model at the center of the task space.

\begin{figure}[]
	\centering
    \includegraphics[width=.40\textwidth]{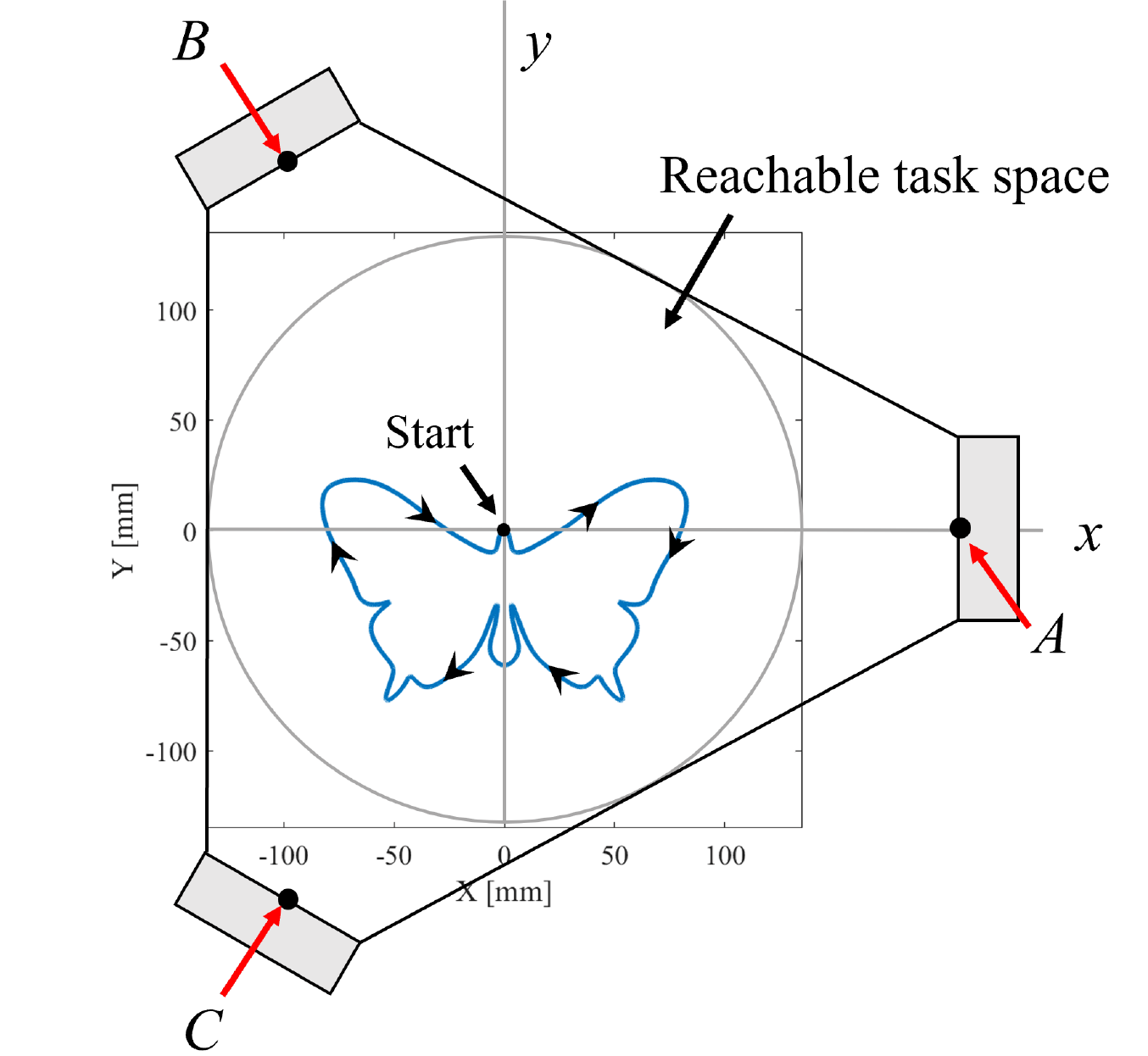}
    \caption{Trajectory of a butterfly used for simulations overlaid on the task space of the delta 3D printer. The butterfly spans $x \in [-82,82]$ mm, $y \in [-77,23]$ mm with a maximum motion speed of $150$ mm/s and a maximum acceleration of $20$ m/s$^2$.}
    \label{fig:simulation_trajectory}
\end{figure}

The simulations are conducted using the trajectory of a butterfly shown in Fig. \ref{fig:simulation_trajectory}, which spans $x \in [-83,83]$ mm, $y \in [-77,23]$ mm with a maximum speed of $150$ mm/s and a maximum acceleration of $20$ m/s$^2$. The trajectory lasts 5 seconds with a sample time of 1 ms (i.e., it contains 5,000 points). The LSR of $\mathbf{G}(s)$ is computed using the known trajectory points and used to simulate the system's response. Parameter-varying compensation for all points (controllers (a) and (b)) is implemented by computing a point-by-point LSR matrix for each window. In other words, we compute the transfer function at each point in the window and compute its impulse response, which becomes the time-shifted columns of the LSR matrix (see Appendix A). For the single point compensation (controllers (c)-(e)), we compute the transfer function and impulse response for the median point in the window, whose time-shifted impulse response is repeated to construct the LSR matrix for each window. All controllers use B-spline basis functions of degree $m = 5$, a window size $L_{C} = 196$ points, number of B-spline coefficients $n = 44$, and number of updated coefficients $n_{up} = 22$. The window size is determined by the amount of time required for the impulse response of a transfer function (IIR filter) to settle close to zero (see \cite{duan2018}). Since the dynamics vary for the delta 3D printer, we construct a grid of positions in the reachable workspace that are 5 mm apart, compute the impulse response for the transfer function in each position, and use the worst-case settling time to determine the $L_{C}$ parameter for all windows. The number of B-spline coefficients is computed from the window length as described in \cite{duan2018}.

\begin{figure}[]
	\centering
    \includegraphics[width=.48\textwidth]{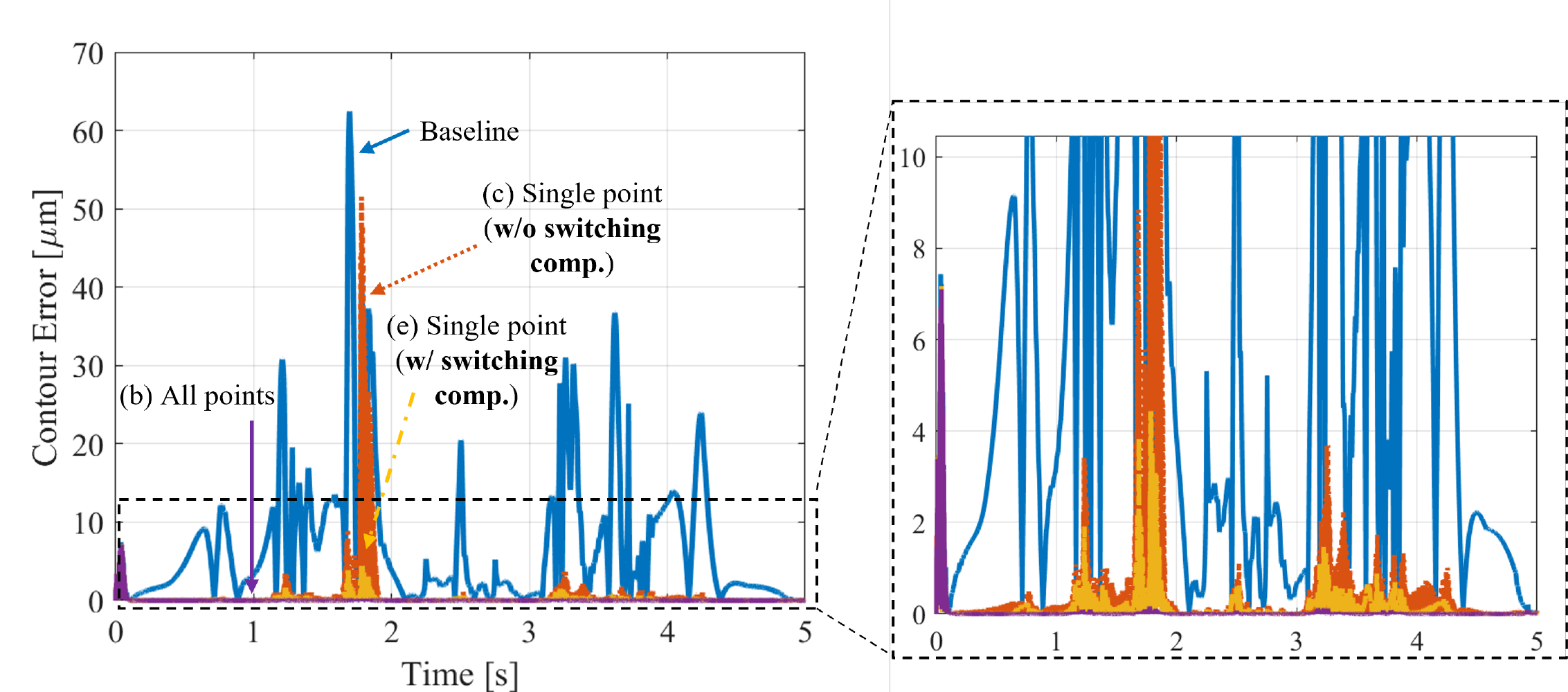}
    \caption{Contour error of the modified trajectories generated by the baseline controller (solid blue line) and controllers (b) using all trajectory points (solid purple line), (c) a single point without switching compensation (dotted red line), (e) and a single point with switching compensation (dash-dotted yellow line).}
    \label{fig:contour_error}
\end{figure}

\begin{table*}[t]
\centering
\caption{Simulation results comparing computation time and accuracy of different controllers for generating modified butterfly trajectory}
\centering
\begin{tabular}{@{}lccc@{}}
\toprule
                                                    & Computation Time & RMS Contour Error & \% Improvement from Baseline \\ \midrule
(*) Baseline LTI, standard FBS controller               & 0.044 s         & 11.13 $\mu$m        & --  \\
(a) Matrix TFs, all points, pseudoinverse               & 820.06 s         & 0.38 $\mu$m        & 96.6\%                                \\
(b) Parameterized TFs, all points, pseudoinverse      & 226.30 s          & 0.38 $\mu$m        & 96.6\%                                \\
(c) Parameterized TFs, single point, pseudoinverse    & 9.86 s           & 3.21 $\mu$m       & 71.1\%                                 \\
(d) Same as above with switching compensation & 13.46 s           & 0.53 $\mu$m       & 95.3\%                                  \\
(e) Same as above with QR factorization        & 9.65 s           & 0.53 $\mu$m       & 95.3\%                                \\ \bottomrule
\end{tabular}
\label{tbl:simulation_results}
\end{table*}
The simulations were run in Matlab (version R2022a) on a 64-bit Microsoft Surface Book with an Intel Core i5-6300U CPU processor and 8 GB of RAM. The computation time of the entire modified trajectory and root-mean-square (RMS) of the contour error across the trajectory points for each controller are reported in Table \ref{tbl:simulation_results}. The percent difference of the RMS contour error compared to the baseline controller simulation is also reported and the contour error comparison is shown in more detail in Fig. \ref{fig:contour_error}. The RMS contour error of the baseline controller is 11.1 $\mu$m and the trajectory is computed in  45 ms because the filtering and inversion is completed offline. The RMS error of the exact LPV model is almost 30 times less at 0.4 $\mu$m. However, note that the computation of the matrix model online is much larger (820 s), which is also about 4 times greater than the computation time of the parameterized model (226 s) without any change in the RMS error. When we compute the model for a single point in each window, we can reduce the computation time by about 20 times (to 10 s) but at a cost of about 10x increase in RMS contour error (3.2 $\mu$m). The accuracy is improved to only be about 1.3x worse than the exact model (about 0.5 $\mu$m) when the switching compensation is implemented. 

In Fig. \ref{fig:contour_error}, there is a spike in the contour error of controller (c) (no switching compensation) around 2 seconds into the motion. The spike represents a difficult portion of the trajectory--the bottom right of the butterfly wing--between which a switch in models occurs for controllers (c)-(e). Here, the points are on the far side of carriage $B$'s line of action (see Figs. \ref{fig:delta_overhead_view} and \ref{fig:simulation_trajectory}) which, as discussed in \cite{edoimioyaTASE2022} and in the following subsection, is prone to larger dynamic variation. (The symmetric points on the bottom left side of the butterfly are not on the far side of carriage $C$'s line of action and, thus, have less dynamic variation). Note that the errors increase for both single point controllers but are exacerbated by the controller without switching compensation. Situations like this one illustrate the utility of using switching compensation for changing models. Finally, note that the computation time reduces by 28\% using QR factorization instead of the pseudoinversion when the controller has switching compensation (from about 13.5 to 9.7 s). Overall, the accuracy of the single point approach is worse than using all points, but the overall accuracy improvement is acceptable given the 23x decrease in computation time compared to using the parameterized exact controller (b), which would be challenging to implement on hardware in real-time. In the following subsection, our experiments on the delta 3D printer show that our proposed controller results in significant accuracy improvements compared to the baseline.

\subsection{Experimental validation} \label{subsec:experiments}

In this subsection, we discuss the results of printing a standard \enquote{calibration cube}\footnote{The standard XYZ calibration cube used in this paper can be found at: \url{https://www.thingiverse.com/thing:1278865}} on the delta 3D printer of Fig. \ref{fig:delta_schematic} using two control strategies: the baseline controller and our proposed controller (case (e) above). Our aim is to demonstrate the utility of our contributions by comparing: (a) the visual quality of parts printed with our controller and the baseline controller at different positions and (b) acceleration amplitudes of the carriages during the execution of each print to understand the effects of our proposed controller.

\begin{figure*}[]
	\centering
    \includegraphics[width=.99\textwidth]{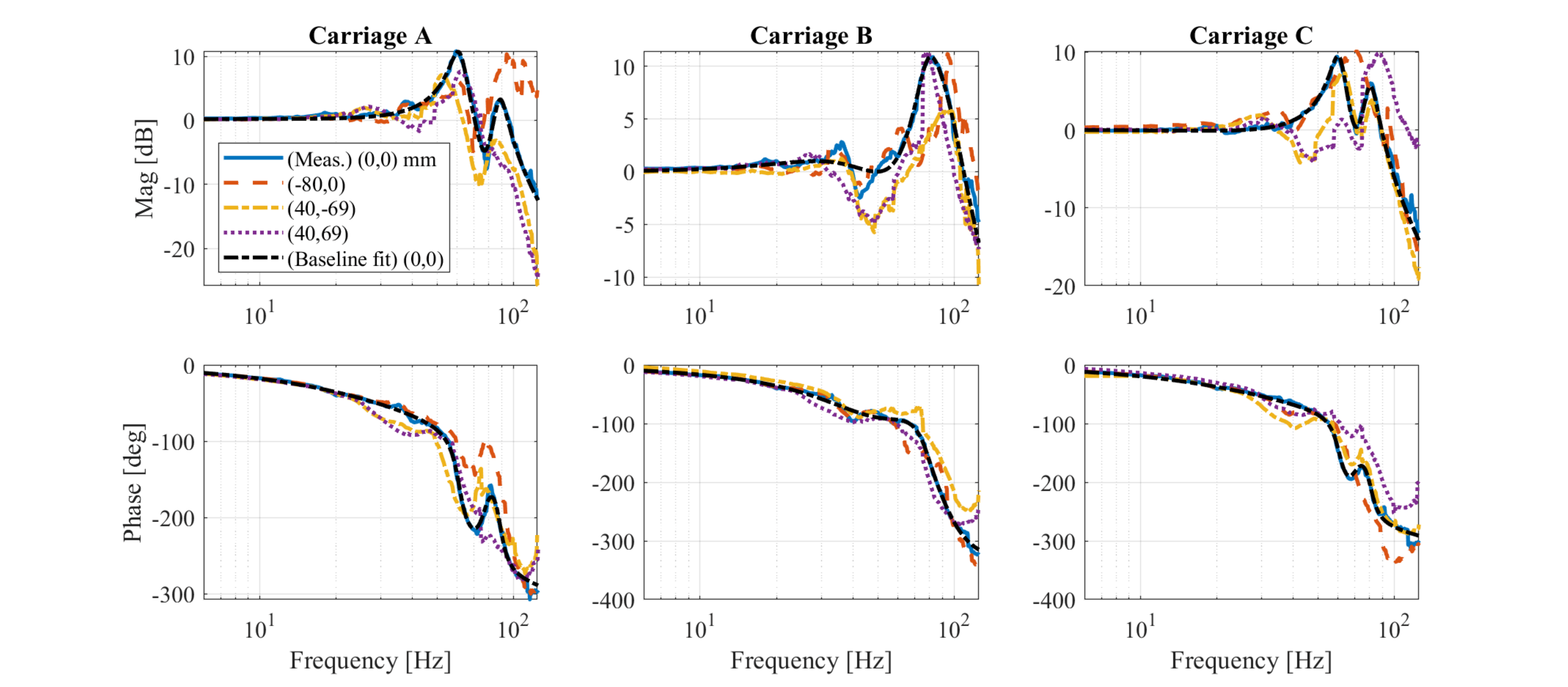}
    \caption{Measured frequency response functions of the carriage position dynamics at $(x,y) = (0,0)$ mm (blue solid lines), $(-80,0)$ mm (red dashed lines), $(40,-69)$ mm (yellow dash-dotted lines), and $(40,69)$ mm (indigo dotted lines) of the Monoprice Delta Pro 3D printer. The black dashed lines indicate the fitted transfer functions of the baseline model (at $(x,y) = (0,0)$ mm) that is used for the baseline controller.}
    \label{fig:baseline_FRFs}
\end{figure*}

For the experiments, we locate the center of the calibration cube at the following positions: $(x,y) =$ $(0,0)$, $(-80,0)$, $(40,-69)$, and $(40,69)$ mm. Each position, except the origin, is chosen to target each carriage independently; they are located 80 mm from the origin along the far side of the respective carriage's line-of-action (see Fig. \ref{fig:delta_overhead_view}). As shown in Figs. \ref{fig:delta_overhead_view} and \ref{fig:baseline_FRFs}, the position $(-80,0)$ mm primarily tests variation in carriage $A$'s dynamics, $(40,-69)$ mm primarily tests variation in carriage $B$'s, and $(40,69)$ mm primarily tests variation in carriage $C$'s \cite{edoimioyaTASE2022}. We set the maximum speed and acceleration at 150 mm/s and 20 m/s$^2$, respectively. Both the baseline and proposed controllers are implemented in Matlab Simulink, which sends motion commands through a dSPACE MicroLabBox to Pololu DRV8825 stepper motor drivers to operate the stepper motors on the delta 3D printer. For the baseline controller, we fit the measured transfer functions at $(0,0)$ mm (see Fig. \ref{fig:baseline_FRFs}) as the LTI model for the standard FBS approach. The proposed controller uses the LPV model and FBS implementation described in Secs. \ref{sec:dynamic_model} and \ref{sec:proposed_controller}, respectively. For comparison, we also print the calibration cube at the same positions without vibration compensation.

\begin{figure*}
    \centering
    \includegraphics[width=0.8\textwidth]{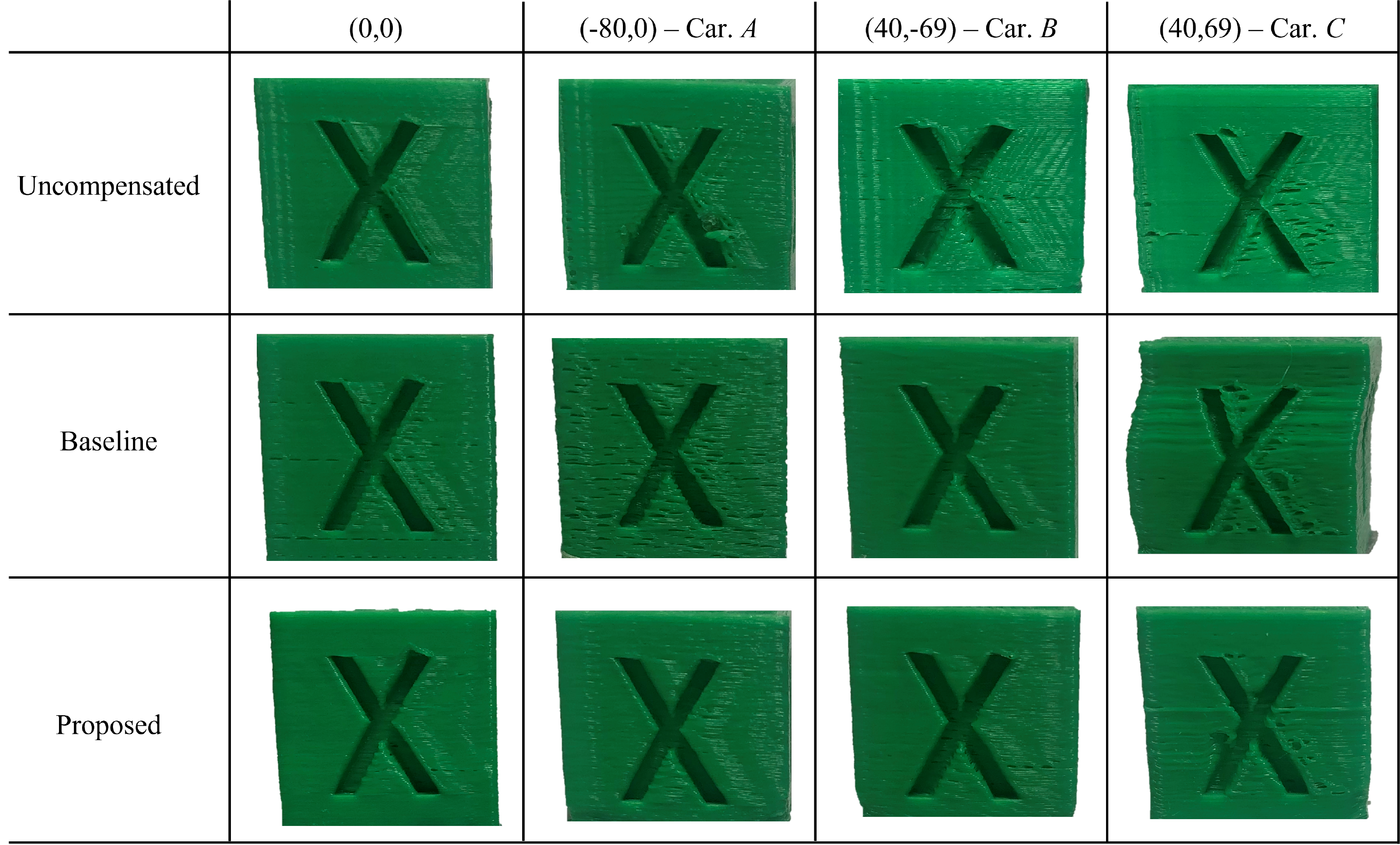}
    \caption{X-axis face of calibration cubes fabricated with the baseline and proposed controllers centered at different positions that target different carriages.}
    \label{fig:X_face_fbs_prints}
\end{figure*}

\begin{figure*}
    \centering
    \includegraphics[width=0.8\textwidth]{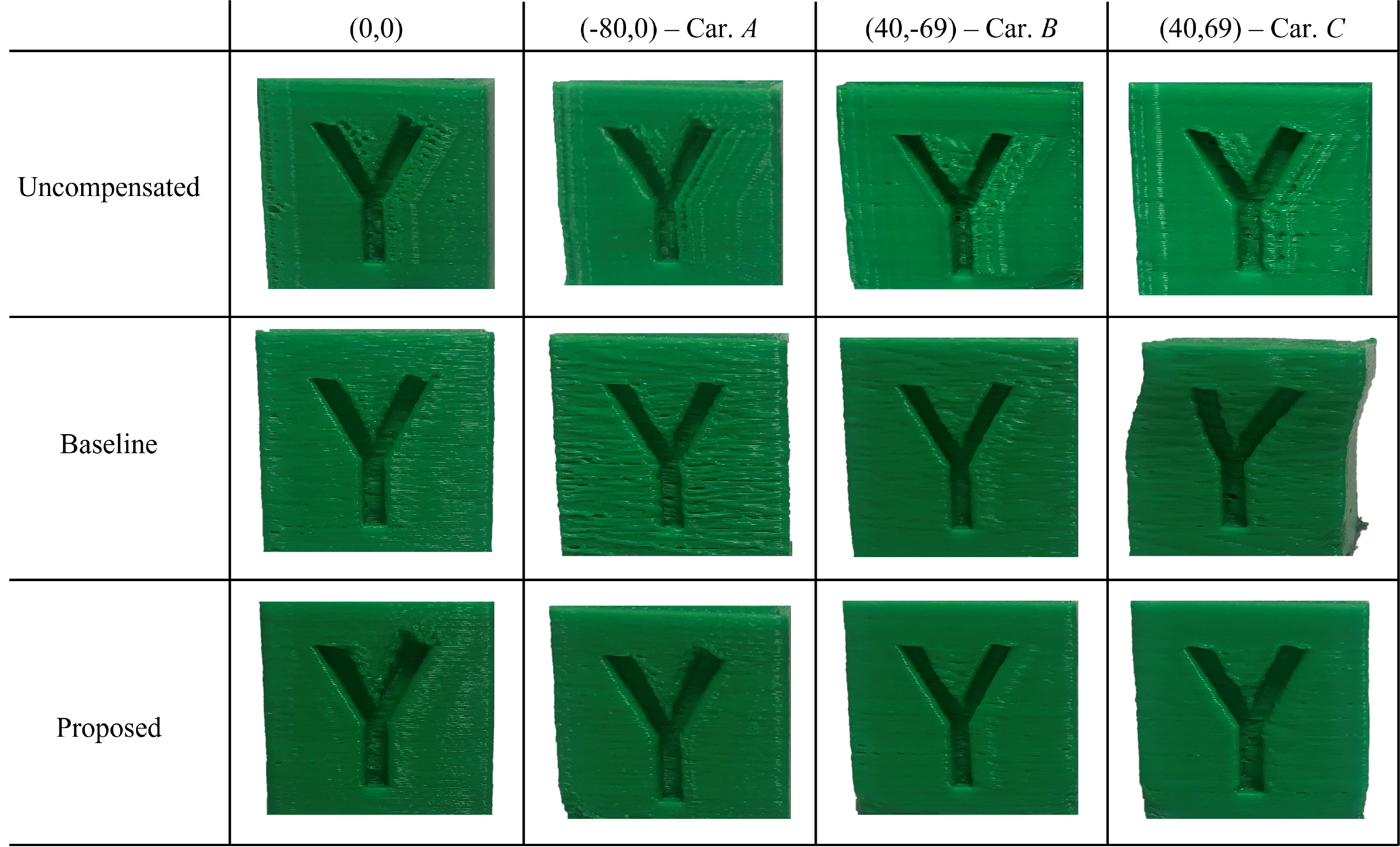}
    \caption{Y-axis face of calibration cubes fabricated with the baseline and proposed controllers centered at different positions that target different carriages.}
    \label{fig:Y_face_fbs_prints}
\end{figure*}

Figures \ref{fig:X_face_fbs_prints} and \ref{fig:Y_face_fbs_prints} show images of the X and Y faces of the calibration cube, respectively, manufactured at the different positions. A visual inspection of the parts reveals the following observations:

\begin{enumerate}
    \item In the uncompensated parts, there are vibration marks at the edges where there is a change of direction, which are largely eliminated with FBS compensation.
    \item The quality of the parts printed at $(0,0)$ mm are similar for both the baseline and proposed controllers.
    \item The surface quality of the part printed at $(-80,0)$ mm with the baseline controller is worse than the quality of the part printed with the proposed controller.
    \item The quality of the parts printed at $(40,-69)$ mm are similar for both controllers.
    \item The part printed at $(40,69)$ mm with the baseline controller drifts from its starting position in the middle of the print, while the part printed with the proposed controller stays aligned.
    \item The quality of the parts printed with the proposed controller are always either similar to or better than the parts printed with the baseline controller.
\end{enumerate}

\begin{figure*}
    \centering
    \includegraphics[width=0.99\textwidth]{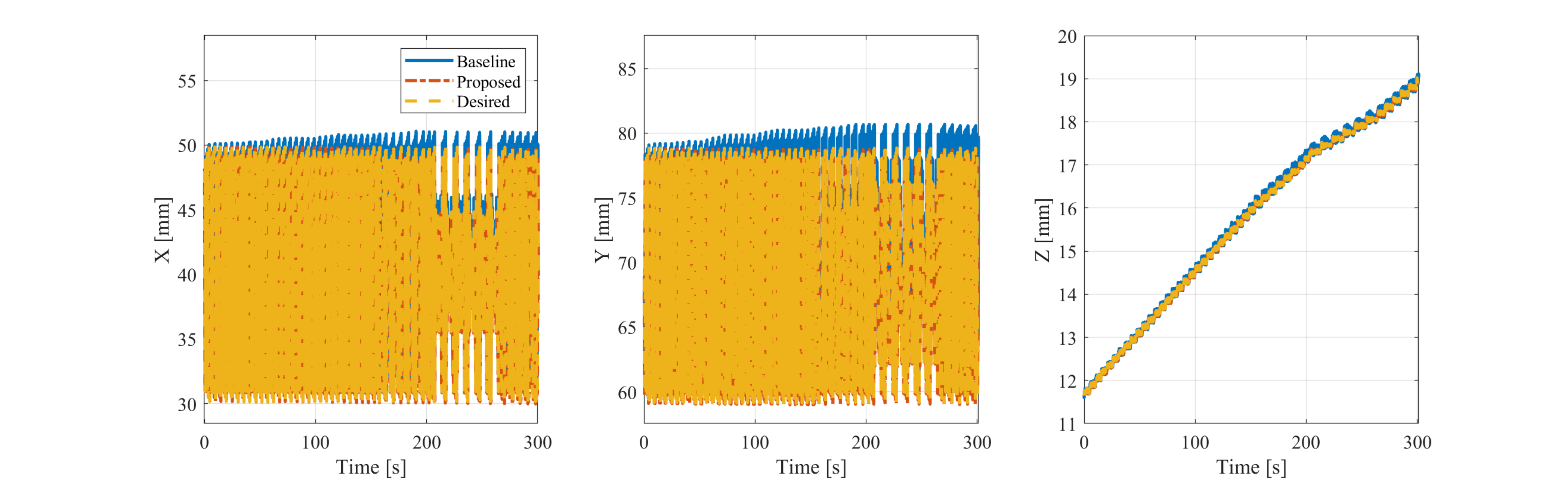}
    \caption{Commanded motion from the baseline (blue solid line) and proposed (red dash-dot line) controller during the drifting motion for the baseline controller while fabricating the part at $(x,y) = (40,69)$ mm (triggering carriage $C$). Note that the baseline model commands increasing deviations from the nominal position (yellow dashed line), which leads to the drifting part in Figs. \ref{fig:X_face_fbs_prints} and \ref{fig:Y_face_fbs_prints}. The baseline controller creates the drifting commands because of the differences between the baseline frequency response function and the actual frequency response function at $(40,69)$ mm.}
    \label{fig:layer_shifting_XYZ}
\end{figure*}

Observation 2 is expected since the baseline controller performs optimally at $(0,0)$ mm and observation 3 is expected due to the model mismatch. However, observation 4 appears to be an anomaly. A closer look at carriage $B$'s FRFs in Fig. \ref{fig:baseline_FRFs} reveals that the measured FRFs from $(0,0)$ and $(40,-69)$ mm have similar resonance frequencies. Also note that carriages $A$ and $C$ have measured FRFs from $(0,0)$ and $(40,-69)$ mm that also have similar resonance frequencies. Hence, the baseline controller is able to adequately compensate vibrations while printing at $(40,-69)$ mm. The drifting signal in observation 5 is due to the baseline controller overcompensating for the fast changes in acceleration on the top half of the Y-face of the cube. Note that the bottom half of the Y-face only has one indentation, while the top half has two indentations in succession, which increases the high frequency content of the acceleration profile. Figure \ref{fig:layer_shifting_XYZ} shows the modified motion commands of the baseline and proposed controllers in this region of the print, which shows that the commanded motion of the baseline controller drifts from the desired command while the  proposed controller does not.  Overcompensation occurs because the baseline model for carriage $C$ (at $(0,0)$ mm in Fig. \ref{fig:baseline_FRFs}) shows that the amplitude of high frequency content is reduced. Hence, the baseline controller attempts to increase the input of the high frequency commands to achieve the desired motion. However, we know from the measured FRF at $(40,69)$ that the command does not need to be amplified. Thus, the proposed controller, with more accurate dynamics, can compensate correctly.

\begin{table}[] 
\centering
\caption{Root mean square (RMS) acceleration of carriages during print of calibration cube}
\begin{tabular}{@{}cccc@{}}
\toprule
                       & Baseline [$\text{m/s}^2$] & Proposed [$\text{m/s}^2$]& Desired [$\text{m/s}^2$] \\ \midrule
$(-80,0)$ -- Car. $A$  & 3.73 (+0.38)              & 3.24 (-0.11) & 3.35                   \\
$(40,-69)$ -- Car. $B$ & 4.04 (+0.08)          & 3.95 (-0.01) & 3.96  \\
$(40,69)$ -- Car. $C$  & 4.16 (+0.35)              & 3.82 (+0.01) & 3.81                   \\ \bottomrule
\end{tabular}
\label{tbl:rms_acceleration}
\end{table}

\begin{table}[]
\centering
\caption{Maximum acceleration of carriages during print of calibration cube}
\begin{tabular}{@{}cccc@{}}
\toprule
                           & Baseline [$\text{m/s}^2$]& Proposed [$\text{m/s}^2$] & Desired [$\text{m/s}^2$]\\ \midrule
$(-80,0)$ -- Car. $A$  & 22.46 (+4.57)              & 17.04 (-0.85)  & 17.89                    \\
$(40,-69)$ -- Car. $B$ & 24.67 (+0.81)    & 24.05 (+0.19)  & 23.86                    \\
$(40,69)$ -- Car. $C$  & 25.53 (+1.67)  & 23.68 (-0.18)  & 23.86     \\ \bottomrule
\end{tabular}
\label{tbl:max_acceleration}
\end{table}
To quantify the reduction of vibration-induced acceleration, we measure the acceleration of the carriages during each print using the vertical ($z$-) axis of an ADXL335 3-axis accelerometer from Sparkfun Electronics and compare the acceleration for both controllers to the acceleration of the desired trajectory. Tables \ref{tbl:rms_acceleration} and \ref{tbl:max_acceleration} give the RMS and maximum values of the carriage acceleration, respectively, during the top half of the calibration cube print. In absolute terms, the proposed controller accelerations are closer to desired acceleration in all cases, illustrating reduction in vibration errors. The maximum difference of deviation reduction of the proposed controller compared to the baseline controller is 8.9\% for carriage $C$ at $(40,69)$ in the RMS acceleration and 20.8\% for carriage $A$ at $(-80,0)$ in the maximum acceleration. Following the result from observation 4, we note that the least deviation from the desired acceleration occurs for carriage $B$ at $(40,-69)$ for both RMS and maximum acceleration. 

\section{CONCLUSIONS}\label{sec:conclusions}

This paper proposes practical techniques to enable real-time, accurate vibration compensation on the prismatic-joint delta 3D printer. Previous work on improving accuracy of delta manipulators has focused on servo motor actuated machines and relies on sensor measurements and feedback control. For most delta 3D printers, feedback sensors are not available so we must employ feedforward control with an accurate model. To achieve this objective, we aimed to use an accurate LPV model of the delta 3D printer we recently proposed in \cite{edoimioyaTASE2022} with the model-inversion based FBS approach. However, the need to recompute the model and controller at each new configuration during real-time control is computationally challenging. Therefore, we propose the following to decrease the computational burden: (1) parameterization and pre-filtering of portions of the model for fast online operations, (2) computation of the model at sampled points along the trajectory (while preserving continuity of the controller's predictions when the model changes), and (3) utilization of matrix methods that yield faster matrix inversion.

Simulations are used to assess the trade off between computation time and accuracy. We report that the techniques presented in this paper result in a 23x reduction in computation time from the exact parameter varying controller which re-computes the model/controller at every point. Thus, our approximations save significant computational effort while only increasing contour errors by about 1.3x compared to the exact controller. Images of parts from our experiments also show an overall improvement in the quality of parts printed at different locations using the proposed controller compared to using a baseline controller optimized for the center of the workspace. Furthermore, acceleration measurements during printing show more than 20\% reduction of vibration-induced accelerations for the proposed controller when compared to the baseline. This work shows that we can take advantage of the high speed motion of the delta 3D printer (compared to traditional 3D printers) and apply feedforward controllers like FBS to maintain accuracy during vibration-prone motion. This paper's contributions bring us one step closer to the vision of high speed and high quality additive manufacturing.

\section*{APPENDIX}

\subsection{Lifted system representation of a digital filter}

As discussed in \cite{ramani2017}, consider digital filter $p$, input signal $u$, and output signal $y$ defined as:
\begin{align}
    p = & \{p_{-2}\hspace{0.5em}p_{-1}\hspace{0.5em}p_{0}\hspace{0.5em}p_{1}\hspace{0.5em}p_{2}\} \label{eq:lsr_filter} \\
    u = & \{u_{0}\hspace{0.5em} u_{1}\hspace{0.5em}u_{2}\} \\
    y = & \{y_{0}\hspace{0.5em} y_{1}\hspace{0.5em} y{1}\} \label{eq:lsr_output}
\end{align}
Signals $y$ and $u$ and filter $p$ are related by the convolution operator as follows:
\begin{equation}
    y = u*p
    \label{eq:convolution}
\end{equation}
From Eqs. \ref{eq:lsr_filter}-\ref{eq:convolution},
\begin{align}
    y_{0} = & p_{0}u_{0} + p_{-1}u_{1} + p_{-2}u_{2} \\
    y_{1} = & p_{1}u_{0} + p_{0}u_{1} + p_{-1}u_{2} \\
    y_{2} = & p_{2}u_{0} + p_{1}u_{1} + p_{0}u_{2}
\end{align}
This can be expressed in matrix form as
\begin{equation}
    \begin{bmatrix}
       y_{0} \\
       y_{1} \\
       y_{2}
    \end{bmatrix}
    =
    \begin{bmatrix}
       p_{0} & p_{-1} & p_{-2} \\
       p_{1} & p_{0} & p_{-1} \\
       p_{2} & p_{1} & p_{0}
    \end{bmatrix}
    \begin{bmatrix}
       u_{0} \\
       u_{1} \\
       u_{2}
    \end{bmatrix}
    \label{eq:lsr_matrix}
\end{equation}
Note that the main diagonal element ($p_{0}$) represents the influence of the current input on the current output; the first upper diagonal element ($p_{-1}$) represents the influence of the succeeding input on the current output and the second upper diagonal element ($p_{-2}$)  represents the influence of the second succeeding input on the current output. Similarly, the first ($p_{1}$) and second lower ($p_{2}$) elements represent the influence of the first and second preceding inputs on the current output, respectively. Hence, the discrete time transform of $p$ obtained from Eq. \ref{eq:lsr_matrix} is given by
\begin{equation}
    p_{2}z^{-2} + p_{1}z^{-1} + p_{0}z^{0} + p_{-1}z^{1} + p_{-2}z^{2}
\end{equation}
which is in accordance with the time-domain definition given in Eqs. \ref{eq:lsr_filter}-\ref{eq:lsr_output}.

\section*{ACKNOWLEDGMENT}

This work was supported in part by the National Science Foundation [grant numbers 2054715 and DGE 1256260] and a Michigan Space Grant Consortium (MSGC) graduate fellowship from the National Aeronautics and Space Administration (NASA), under award number 80NSSC20M0124. A company founded by C.E. Okwudire holds a commercial license for the filtered B-splines (FBS) algorithm.

\addtolength{\textheight}{-4cm}   






\end{document}